# Magnetoelectric effect induced by the delocalized $^{93m}$Nb state

[1,2,3]Yao Cheng (程曜), [4]Yuan-Long Liu (刘源龙), [5,6]Chi-Hao Lee (李志浩), [4]Ben-Li Young (杨本立)

[1] Center for Nano and Micro Mechanics, Tsinghua University, Beijing, Haidian, 100084, China

[2] Institute of Advanced Study, Nanchang University, Nanchang, 330031, China

[3*]Department of Engineering Physics, Tsinghua University, Beijing, Haidian, 100084, China

[4]Department of Electrophysics, National Chiao Tong University, Hsinchu, 300, Taiwan

[5]Department of Engineering and System Science, National Tsinghua University, Hsinchu, 300, Taiwan

[6]National Synchrotron Radiation Research Center, Hsinchu, 300, Taiwan

**Abstract**

We report a temperature- and density-dependent decay of the $^{93m}$Nb nuclear excitation and give a minimal interpretation on the underlying physics. This anomaly indicated nuclear resonant absorption as well as delocalization of the long-lived Mössbauer state in the crystal. A nonlinear magnetoelectric response, on low-frequency drive current, showed up in the bulk metal of a high-purity niobium crystal and then disappeared with vanishing benchmarks of delocalized nuclear excitation. Several nonlinear resonant peaks, on the order of several hundreds of Hz, grew up with the applied magnetic field. The central frequencies of these peaks decreased with temperature.

**Introduction:**

When a laser shines on an atom, with a frequency near the atom's resonant transition, the laser's intensity tunes the transition rate to give the Rabi oscillation [1]. However, no γ-ray laser has yet been reported [2]. The lifetime for the nuclear transition of γ decay cannot be tuned, making it a constant standard. We report a pronounced increase in the rate of γ decay for $^{93m}$Nb from its internal conversion [3], of which the half-life shall be 16.13 years [4]. Furthermore, this anomalous decay became faster at low temperature. When resonant absorption occurs in a crystal consisting of identical nuclei, the nuclear excitation is no longer fixed at one particular nucleus but spreads out [5]. Some particular impurities in the crystal become leakage channels for the delocalized nuclear excitation, resulting in a temperature- and density-dependent decay.

The diffusive transport of electrons in a bulk metal is purely resistive, with voltage having no phase shift in response to an ac drive current, particularly at low frequency. Furthermore, no transverse response from an applied field will appear in a bulk metal without any significant orbital magnetization [6]. We also report a nonlinear transverse magnetoelectric (ME) response on low-frequency drive current in a high-purity niobium crystal, which is associated with the anomalous $^{93m}$Nb decay. The ME response showed up not only in the transverse term but also in the longitudinal term, both of which are symmetrical to the positive and the negative applied magnetic field. Plenty of nonlinear resonant peaks between 200 and 1000 Hz were observed. We shall demonstrate that these ME effects are due to the delocalization of the $^{93m}$Nb excitation and take this exotic phenomenon as evidence for the long-lived Mössbauer effects [7].

The magnetoresistance of a bulk metal, induced by the Lorentz force, increases with the quadrate of the field strength, and is insignificant in nonmagnetic metals [8]. Obtaining a giant magnetoresistance requires a composition of ferromagnetic and non-magnetic thin layers [9]. Recently, the revival of ME effects [10] has blossomed into a vital research field connected with topological insulators [11]. Interestingly enough, most of the reported 3D topological "insulators" are actually not insulators but helical metals [12], *i.e.*, a metallic bulk with helical surface states, of which the ME response is no longer quantized. The ME effect of the antiferromagnetic chromium sesquioxide, $Cr_2O_3$, is weak [10]. The coupling between ferromagnetism and ferroelectrics may induce a gigantic ME from the multiferroics, and thus simultaneous magnetic and electric orderings must be pursued [13]. Our sample is neither ferromagnetic nor ferroelectric.

---

* Contract between the main author, YC, and Tsinghua University was interrupted for more than one month during this work. This work has been done independently from the department of engineering physics, Tsinghua University, Beijing.

In a two-dimensional electron gas under a magnetic field, a topological edge state is free from back scattering, such that the skipping motion of electrons is perfectly quantized. This is known as the integer quantum Hall effect (IQHE) [6,11]. A magnetic moment of circulating Hall current appears, when the electric field on the surface is applied. The ME effects at the IQHE-plateau transitions behave as if they are nonlinear, *i.e.* the E-field induced magnetic moment depends on the magnetic field generated by the magnetic moment itself. However, the axion electrodynamics remains linear on the quantum plateaus, when the field change is sufficiently small so as not to cross the quantum steps.

Niobium is a special nuclear species, which has only one natural isotope and a long-lived Mössbauer state. The first low-lying states of niobium, rhodium, yttrium, and scandium have half-lives of 16 yr, 56 min, 16 s and 0.3 s [4], respectively. All of these nuclides have 100% natural abundance. According to the uncertainty principle, the long lifetime gives extremely narrow linewidths for the nuclear states, ranging from $10^{-24}$ eV to $10^{-15}$ eV. Conventional wisdom holds that resonant absorption of the long-lived Mössbauer photon is impossible due to the broadening of their linewidths, to varying degrees far beyond what is intrinsic [14]. Plenty of exotic phenomena with scandium and rhodium have long been observed for years [7,15]. Major long-lived Mössbauer effects that have been reported include: Rabi oscillation of x rays as a function of temperature and excitation density; anomalous x rays indicating biphoton cascade transitions; anisotropic nuclear susceptibility depending on sample geometry at room temperature; and a quantum phase transition showing collapse and revival. None of the varied interpretations, across several different disciplines, has gained widespread acceptance. Wanting to resolve this long-standing problem has motivated us to search for evidence in low-temperature from $^{93m}$Nb which has a sufficiently long half-life.

We here briefly introduce the subtle physics of the long-lived Mössbauer effects. The nuclear resonant absorption of multipolar transitions is not mediated by a single photon but by two entangled photons, of which wavelength matches the lattice constant [7]. Photo-electric effects are forbidden by the multipolar features of the EM field [5]. Once the nuclear resonant absorption occurs, excitation spreads out to the identical nuclei of a crystal like a delocalized Mott-Wannier exciton. These delocalized nuclear excitons do not consist of pairs of electrons and holes but of pairs of protons and proton-holes propagating with an entangled photon pair, as the so-called nuclear spin-density wave (NSDW) [7]. NSDW dressing [1] on nuclei is confined in crystal to become massive. Due to the light mass on the eV order, NSDW of a micron size has a huge magneton to give the observable nuclear susceptibility at room temperature. The translation symmetry of Bloch electrons is thus broken by the magnetic lattice of NSDW. When the excitation density of $^{103m}$Rh exceeds $10^{12}$ cm$^{-3}$, we observed a quantum phase transition showing collapse and revival. This massive NSDW most-likely undergoes Bose-Einstein condensation (BEC) at room temperature, as do entangled microwaves [16]. Due to the long lifetime of $^{93m}$Nb, its vacuum coupling is on the order of meV, which is $10^{-5}$ times the vacuum coupling of $^{103m}$Rh [7]. Low-density Nb NSDW is thus destroyed by thermal fluctuation at room temperature. Nb NSDW survives from thermal fluctuation by means of strong coupling [1] with excitation density beyond $10^{12}$ cm$^{-3}$, as reported here. We anticipate a direct NSDW BEC at room temperature, because the mass of $^{93m}$NB NSDW shall be lighter than the mass of $^{103m}$Rh NSDW.

**Experiment:**

We prepared a "sample A", consisting of a single-crystal Nb [(110)-oriented oval plate of 1.2mm × 12mm × 13mm], by neutron irradiation with 2×10$^{12}$ n/cm$^2$s for 5 hours, in the reactor at Tsinghua University, Hsinchu, in 2008, while keeping, as a reference, an identical "sample B" without irradiation. Most of the thermal neutrons were removed by a Cd envelope so that only fast neutrons with a flux of 10$^{11}$ n/cm$^2$s passed through the sample. Simply using the Nb Kα from the internal conversion of the $^{93m}$Nb state, we estimated the neutron irradiation of sample A to have an excitation density of 2×10$^{13}$ cm$^{-3}$. However, we later realized that this estimate is incorrect. Part of the Nb Kα emission was contaminated, having come from $^{94}$Nb, as next detailed. The enhanced decay of sample A at room temperature was first discovered in October, 2010. The increased decay rate at low temperature was then observed in June, 2011. A systematic study of these effects has been done and is reported here.

Three years after the 2008 activation, the ME effects of sample A were unexpectedly discovered, first in February, and then in June, 2011. However, these ME effects disappeared in November, 2011, when the true $^{93m}$Nb excitation density was below 10$^{12}$ cm$^{-3}$. To bring the ME effects back, we irradiated two single crystals again in November, 2011. Sample A was reactivated and irradiated for the second time, while sample B was activated for the first time. The ME effects of sample A were recovered but showed a different behavior from the former observations in June. We conclude that the ME response depends on the $^{93m}$Nb excitation density.

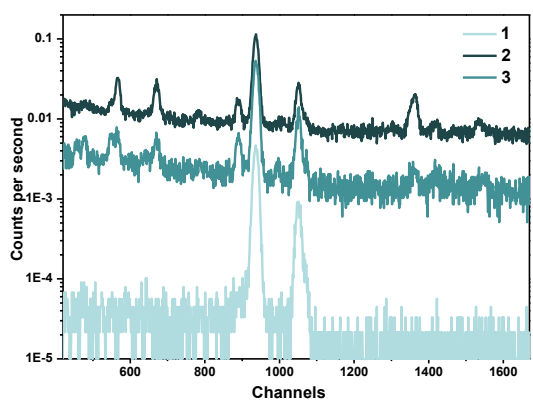

Figure 1: (Color on-line) Spectra of Sample A taken at a different time scheme. Curve 1: before the reactivation by neutron irradiation; Curve 2: one week after the reactivation; Curve 3: three weeks after the reactivation, with two weeks at 4.2 K after spectrum 2 was taken. The numerous peaks from channels 500 to 800 are Hg and W L-lines. The peaks centered at channel 938 and 1053 respectively are Kα and Kβ of Nb. The peaks centered at channels 891 and 997 respectively are Kα and Kβ of Zr. The peak centered at channel 1421 is Kα of In.

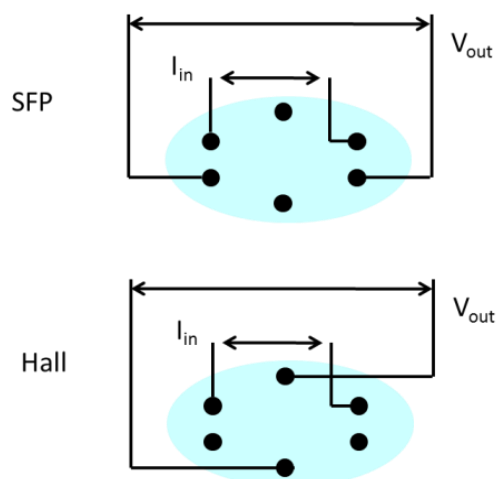

Figure 2: (Color on-line) Configurations of the measurement: the standard four-point (SFP) method and the connection for the Hall effect.

Figure 1 depicts three x-ray spectra of sample A taken at a different time scheme. Spectrum 1 was taken before the reactivation in November 2011, spectrum 2 was taken in one week after the reactivation in November 2011, and spectrum 3 was taken in three weeks after the reactivation in December 2011, *i.e.*, after the ME measurement with the temperature down to 4.2K for two weeks. Spectrum 1 shows only Kα and Kβ emissions from the internal conversion of $^{93m}$Nb decay. All characteristic X-rays from impurities, observed in 2008, had disappeared. The read count rate of Nb Kα in spectrum 1 was 0.068 cps (counts per second), which should truly reflect the excitation density of $^{93m}$Nb. Accordingly, the $^{93m}$Nb density was ~10$^{12}$ cm$^{-3}$. We are unable to give a more precise estimation of the $^{93m}$Nb density, due mainly to the rough ratio of internal conversion in ref. [4]. Spectra 2 and 3, after reactivation, recovered the interesting characteristic

X-rays of impurities observed in 2008, particularly the W L-lines, the Zr K-lines, and the Y K-lines (Y K-lines are insignificant in figure 1 but observable from an activated polycrystalline Nb sample). One might conclude that the W L-lines come from the β decay of $^{182}$Ta for $^{182}$W with a half-life of 114 d, which was created by the neutron capture of $^{181}$Ta impurity in the Nb crystal. However, our monitoring of W L-lines in 2009 revealed a much slower decay, which indicated that another channel was emitting these W x-rays. X-rays of Mercury and Indium, in spectra 2 and 3, disappeared in January 2012. They were due to the contamination from Indium solder and a Cd envelop on the surface, while the Zr K-lines are benchmarks for the reported anomalous decay. The count rates of Nb Kα, in spectra 2 and 3, were respectively 1.58 and 0.76 cps. In the meantime, sample B, at room temperature during the two weeks, showed an enhanced decay of 22%, from 1.13 to 0.89 cps. At low temperature, the anomalous decay of sample A increased from 1.58 to 0.76 cps, in the same two weeks instead of 16 years. The subsequent monitoring of sample A, at room temperature, showed a decay of 13% in two weeks. This relaxed room-temperature decay rate for sample A was even lower than the decay rate of the reference sample B which was at room temperature the entire time. We conclude that the anomalous decay depends on the temperature and the excitation density.

Nb crystal is simply a natural photonic crystal for NSDW [7], in which high photon fields shall be induced at impurity sites. The observation of impurity emissions associated with the enhanced $^{93m}$Nb decay indicates a delocalized NSDW. Below a certain $^{93m}$Nb level near $10^{12}$ cm$^{-3}$, the characteristic X-rays from impurities disappeared along with the reported ME effects. Zr, Y, and W are Nb-crystal impurities, as is the $^{94}$Nb (lifetime $2.03 \times 10^4$ yrs) produced from neutron capture by $^{93}$Nb. The β decay of $^{94}$Nb gives no Nb x-rays but does give Mo x-rays. The misleading Nb Kα, in spectra 2 and 3, are from $^{94}$Nb. The main impurity in a single-crystal sample is Ta, from which x-rays were missing in spectra 2 and 3. The criterion to select leakage channels is simple. The low-lying transitions, e.g. $^{182}$W (E2), $^{90}$Zr (E0), $^{91}$Zr (E2), $^{92}$Zr (E2), $^{94}$Zr (E2), $^{96}$Zr (E0), $^{89}$Y (M4), and $^{94}$Nb (M3) [4], are candidates for NSDW; *i.e.* favoring cascade transitions to absorb or to emit two entangled photons [7]. The off-resonant nuclear excitation induced by NSDW condensate ejects the orbital electrons of impurities to give the characteristic x rays. On one hand, the photon field at impurities is stronger at lower temperature. On the other hand, the virtual excitation of multipolar channels (E2, M4, and M3), averaging over the thermal motion, is enhanced by lowering the temperature. The anomalous decay thus depends on the temperature and the excitation density.

We drilled six holes in sample A, wired the Cu leads by Indium solder, and measured the voltage from an ac drive current by means of a standard four-point configuration (SFP) and a Hall configuration, as depicted in figure 2. When the 5-V voltage from the lock-in output was applied, 79 mA went through the sample. Samples were mounted on a cold head with an upwards normal vector parallel to the applied field in the helium-cooled Dewar. The ac drive of figure 2 confirmed the superconductive transition at 9.3 K. The four-point configuration, in figure 2, gives sample A a resistance of 92 μΩ at room temperature, and a residual resistance of 4.3 μΩ at 10 K.

When a dc drive was applied, the irradiated sample behaved like a normal Nb sample, except that the residual resistance was four-times higher than the residual resistance of the virgin sample without irradiation. The sample at 10K was purely resistive from 2 to 1000 Hz without an applied field. This behavior drastically changed, when the magnetic field was turned on in the normal direction of the sample disc. A pronounced ME response, accompanying a nonlinearity, was found at low frequency, e.g. 10 Hz.

Increasing the field with the SFP configuration, the ME-response phase rotated counterclockwise, changing from capacitive to inductive, as listed in Table 1. Lowering the drive current from 79 mA to 10 mA, at 9 T, the phase also rotated counterclockwise such that the ME response became more inductive, while the amplitude nonlinearly decreased. The ME response could also be tuned by temperature, as shown in Table 1.

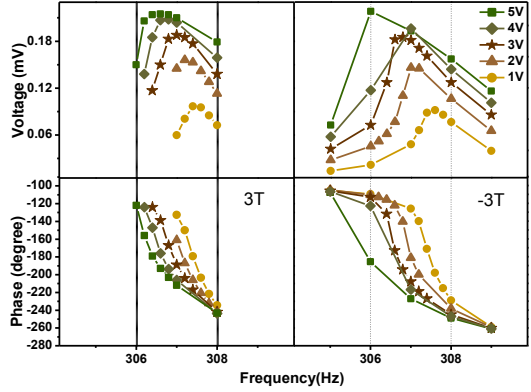
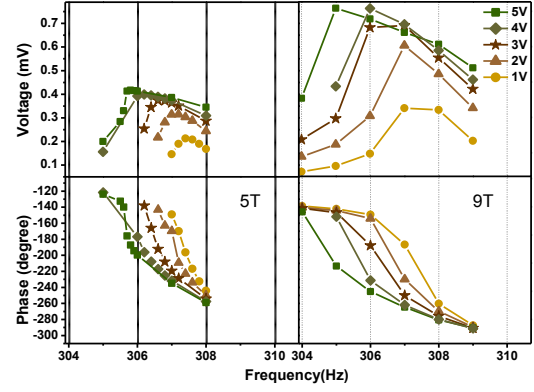

Figure 3A: (Color on-line) Nonlinear response near 306 Hz, in response to current settings (5V for 79 mA), measured with applied fields of 3T and -3T at 4.2 K.

Figure 3B: (Color on-line) Nonlinear response near 306 Hz, in response to current settings (5V for 79 mA), measured with applied fields of 5T and 9T at 4.2 K.

Switching from the SFP configuration to the Hall configuration without applying a B-field at room temperature, the resistance was 14 µΩ, instead of 0 Ω, corresponding to 15% of the 92 µΩ measured using the SFP configuration. This resistance was not a transverse term but still a longitudinal term, due to the asymmetrical position of the wiring holes. We demonstrated this residual longitudinal term also by the superconducting transition with a 0.5-µΩ resistive step at zero field. The transverse ME responses at 5T and 9T, as listed in Table 2, significantly deviated from 15% of the SFP magnitude listed in Table 1. Surprisingly, the phases rotated in a different direction from the field, *i.e.* clockwise instead of counterclockwise, to give a first indication that this observation was something different than a longitudinal term.

Plenty of resonant peaks were discovered between 200 and 1000 Hz. The resonant amplitude and phase depended not only on the drive current in a nonlinear manner, as depicted in figures 3, but also particularly on whether the peak was approached from higher or lower frequency, typically for a nonlinear resonance. Some peaks showed a stronger 3$^{rd}$ harmonic than 2$^{nd}$ harmonic, while others showed a stronger 2$^{nd}$ harmonic than 3$^{rd}$ harmonic. We show, in figures 3, the particular 306-Hz peak, of which the 2$^{nd}$ harmonic is greater than the 3$^{rd}$ harmonic. The phase crossed 180 degrees very quickly at the peaks, which indicated a negative resistance. The structure of the peaks, i.e. frequency, width, amplitude and phase, was different in the two configurations; so it was hard to identify a clear correspondence of peaks between the two configurations. However, the transverse peaks are much greater than 15% of the nearby longitudinal peaks. We conclude accordingly that the transverse ME peaks with Hall configuration are not artifacts due to the asymmetrical wiring positions, but constitute true physics. The ME responses reported here are symmetric to the positive and negative applied field, as depicted in figure 3A. These observations are definitely not of the Hall effect.

The peak frequencies shifted side, from high to low frequency, when the sample was heated from 4.2 to 200 K. Therefore, we believe that the peaks will show up at room temperature, if a strong magnetic field is applied. It mattered not whether the path involved heating or cooling, an interesting frequency jump was observed, repeatable at 4.4 K, which might indicate a phase transition of NSDW.

| Field | 1T | 2T | 5T | 9T |
|---|---|---|---|---|
| @4.2K | 0.30∠-20° | 0.38∠-14° | 0.38∠-8.4° | 0.41∠3.9° |
| @10K | 0.33∠-1.6° | - | - | 0.38∠17° |

Table 1: Longitudinal ME responses (magnitude in µV and ∠phase in degree), using the SFP configuration (fig.2) with a 79-mA ac drive current of 10 Hz, where the residual voltage at 10 K is resistive 0.34 µV at zero applied field.

| Field  | 1T       | 3T       | 5T       | 9T        | -3T      |
|--------|----------|----------|----------|-----------|----------|
| @4.2K  | 0.02∠-25°| 0.05∠-60°| 0.13∠-86°| 0.33∠-102°| 0.07∠-38°|

Table 2: Transverse ME responses (magnitude in μV and ∠phase in degree), using the Hall configuration (fig. 2) with a 79-mA ac drive current of 10 Hz, where the residual voltage at 10 K is resistive 0.04 μV at zero applied field.

**Discussion:**

Anomalous nuclear decay and ME responses depending on the temperature and the $^{93m}$Nb excitation density are two exotic phenomena that emerge at the same time, when the $^{93m}$Nb excitation exceeds a particular threshold. This could have a very favorable impact on the suggestion that the long-lived Mössbauer transition is a means to measure gravitational waves (GW) [17]. The delocalized nuclear excitation, which is our explanation for the anomalous nuclear decay, is necessary for the sensitivity of GW detection [18].

However, the physics community currently does not accept the NSDW concepts; and we are unaware of a good explanation for the present ME results, in terms of standard physics.

Given the unusual nature of the proposed NSDW BEC and the associated GW detection, the main author, YC, thus believes it best to forgo a further discussion at this time, and requests, rather, that the physics community provide comment on the above experimental results and explanations thereof.

**Acknowledgements:**
We thank Cheng-Chung Chi (齐正中) from National Tsinghua Universities Hsinchu, Minn-Tsong Lin (林敏聪) from National Taiwan University for valuable discussions, Shuan-Ya Huang (黄宣雅) for the data preparation and Ken Sasaki for proofreading.

18. Delocalized nuclear excitation gives a collective response on the long-wavelength gravitational waves.